\documentclass{PoS}
\usepackage{graphicx}
\usepackage{epsfig}
\usepackage{subfig}
\usepackage{caption}
\usepackage{multicol}
\usepackage{textcomp}
\usepackage{wrapfig}
\usepackage{enumitem}
\usepackage{adjustbox}

\def \etal{{\em et\ al.},\ }

\def \etal{{\em et\ al.},\ }

\def \aap{Astron. Astrophys.}

\def \aj{Astron. J.}

\def \apj{Astrophys. J.}
\def \apjl{Astrophys. J., Lett.}

\def \mnras{Mon. Not. R. Astron. Soc.}
\def \nat{Nature}

\title{The TeV Morphology of the Interacting Supernova Remnant IC 443}

\ShortTitle{TeV Morphology of IC 443}

\author{\speaker{Brian Humensky}\\
        Columbia University\\
        E-mail: \email{humensky@nevis.columbia.edu}}
\author{for the VERITAS Collaboration\\
        \\
        E-mail: \email{veritas@veritas.sao.arizona.edu}}
\abstract{The middle-aged supernova remnant IC 443 is interacting with molecular gas in its surroundings. \emph{Fermi}-LAT has established that its gamma-ray emission at low energies shows the "pion bump" that is characteristic of hadronic emission. TeV emission was previously established by MAGIC and VERITAS at a site of interaction between the shock front and a molecular cloud. VERITAS has continued to observe IC 443 and can now resolve the emission on few-arcmin scales. We will present results on the emission morphology and discuss possible sources of the emission, including the shell of the remnant and other gaseous structures in the vicinity.}

\FullConference{The 34th International Cosmic Ray Conference,\\
		30 July- 6 August, 2015\\
		The Hague, The Netherlands}

\begin{document}

\section{Introduction}
Supernova remnants (SNRs) appear to meet the energetics requirements to serve as the sources of the Galactic cosmic rays, and are well known as electron accelerators from their synchrotron radiation, which spans from radio to, for some young SNRs, X-rays (e. g., \cite{Koyama1995}). However, it is only in recent years that clear evidence is accumulating of their role in hadronic cosmic-ray acceleration. IC 443 has become one of the key supernova remnants in this story. IC 443 is a middle-aged SNR (age $\sim3-30\ \textrm{kyr}$~\cite{Petre1988,Troja2008,Chevalier1999,Olbert2001}) located towards the Galactic anti-center at a distance of $1.5\ \textrm{kpc}$; it has a diameter of $0.75\textrm{\textdegree}$, corresponding to 20 pc~\cite{Poveda1968,Welsh2003}. It is the result of a core-collapse explosion, and shows clear evidence of interactions with molecular clouds (MCs) through tracers of shocked gas visible at multiple points around the shell (see, for example,~\cite{Seta1998,Rosado2007,Claussen1997,Hewitt2006}). IC 443 has been extensively studied at all wavelengths, and in recent years has been established as a strong gamma-ray source, in both the GeV~\cite{AGILE_IC443,FERMI_IC443} and TeV~\cite{2007ApJ...664L..87A,VER_IC443} bands. Most recently, an updated analysis of \emph{Fermi}-LAT data extending down to $60\ \textrm{MeV}$ shows at low energy the spectral shape expected by gamma-ray production via the decay of neutral pions, the kinematics of whose production creates a spectral signature that is hard to replicate via leptonic mechanisms~\cite{2013Sci...339..807A}. 

VERITAS has continued to observe IC 443 since the initial detection, and now has an exposure of approximately 155 hours livetime. With an angular resolution better than 0.1\textdegree\ (68\% containment) at 1 TeV, VERITAS is able to resolve the emission from IC 443 and investigate its relationship to the morphology of the SNR and the gaseous structures with which it is interacting. Results from these data are discussed below.

\section{Data and Analysis}
VERITAS observations of IC 443 began in February, 2007, and concluded in January, 2015, spanning three epochs in the evolution of the instrument. These data include 52.5 hours taken in the first epoch, prior to the relocation of one telescope to improve the distribution of baselines between the telescopes; 42.4 hours in the second epoch, following the telescope relocation but prior to the camera and trigger upgrades; and finally, an additional 81.9 hours following the upgrades of the camera triggers during 2011-12 and the PMTs in the summer of 2012. In total, 177 hours were taken, resulting in 155 hours of usable data after quality selection and deadtime. This data set is an increase by a factor of 4.5 over the exposure used in~\cite{VER_IC443} for the sky maps, and a factor of 9 for the spectra.

The analysis follows standard procedures (see, for example,~\cite{lsi}), using "moderate" cuts and requiring at least three images passing image quality cuts (and four in the portion of the first epoch for which four telescopes were operational). This leads to an energy threshold (defined as peak of the differential counting rate) of $E_{thr} \sim 240\ \textrm{GeV}$, with spectral reconstruction beginning at $190\ \textrm{GeV}$. The excess map is constructed using an integration radius (top-hat smoothing) of 0.09\textdegree, optimized for a point source, while the spectra are extracted from three regions using integration radii of 0.13\textdegree, as well as from the entire SNR using an integration radius of 0.30\textdegree.

\section{Results and Discussion}
\begin{figure}[h!]
\centering
\adjustbox{trim={.01\width} {.01\height} {0.01\width} {.01\height},clip}{\includegraphics[width=5in]{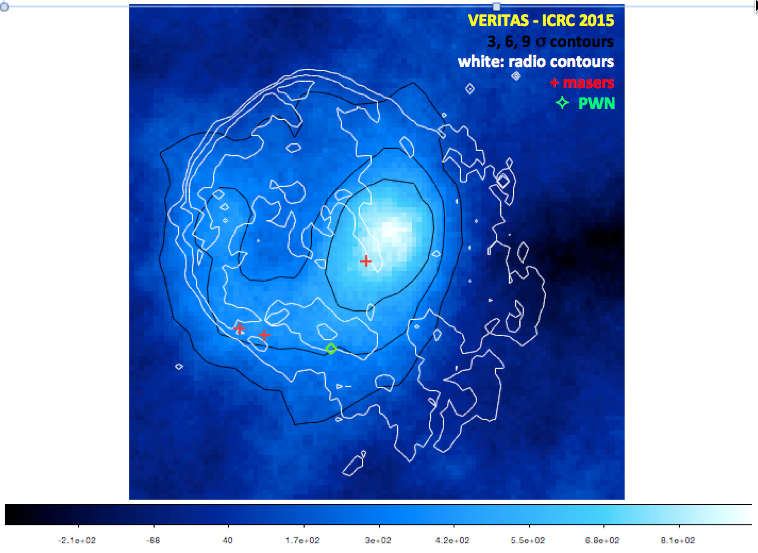}}
\caption{Excess map for the field including IC 443, with the color scale indicating counts integrated within a radius of 0.09\textdegree\ about each point. White contours indicate the radio shell and black contours indicate the significance of the VERITAS observations at the 3, 6, and 9 $\sigma$ levels. Locations of maser emission are marked in red, while the location of a likely pulsar wind nebula is marked in green.}
\label{fig:veritasExcessMap}
\end{figure}

Figure \ref{fig:veritasExcessMap} shows the excess map in IC 443 and the surrounding region, computed using a point-source integration radius. Gamma-ray emission above 200 GeV fills the northern lobe of the SNR, tracing out the SNR/MC interaction regions and the  shell  along much of its length. The emission is strongest where the maser emission is brightest, but the entire remnant appears to be accelerating particles.

\begin{figure}[h!]
\centering
\null\hfill
\subfloat{\includegraphics[width=0.4\textwidth]{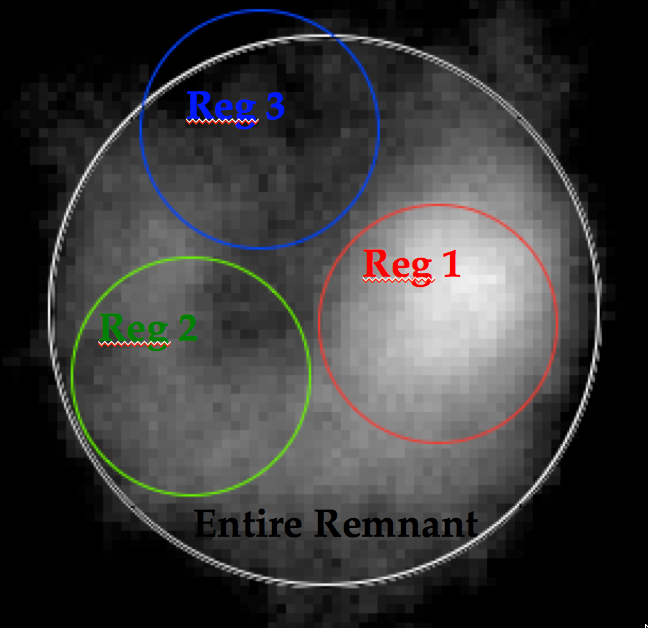}}
\hfill
\subfloat{\includegraphics[width=0.55\textwidth]{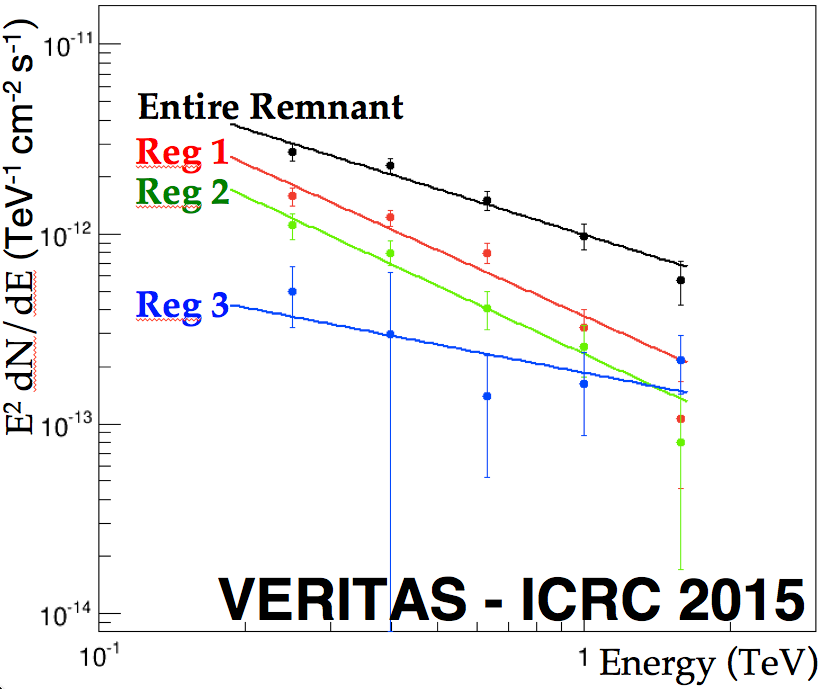}}
\hfill\null
\caption{(left) Excess map, indicating regions from which spectra were extracted. (right) VERITAS spectra for each of three regions and for the entire remnant.}
\label{fig:spectra}
\end{figure}

To study the spectrum of the remnant, we integrate all counts within an 0.3\textdegree\ radius around a point near its center, 06 16 52.8 +22 33 00, as well as selecting out three regions of radius 0.13\textdegree\ that sample different environmental conditions. Those three regions are (1) centered on the brightest maser emission, (2) covering the dim, extended maser emission along the southeast, and (3) in the north, where the shell is interacting with swept-up material and no molecular clouds are observed. The resulting spectra are shown in Figure~\ref{fig:spectra}, and results of power-law fits to the spectra are summarized in Table~\ref{tab:spectra}.  The bin width is 0.2 decades, and optimized for the brighter regions, resulting in relatively poor statistics per bin in the dimmest region. All three regions, as well as the whole SNR, have power-law indices near $\Gamma \sim -3$, consistent with previous studies in this energy range~\cite{2007ApJ...664L..87A,VER_IC443}. This is unusually soft for SNRs but observed in some other SNRs interacting with molecular clouds (eg, W44~\cite{2013Sci...339..807A}). There is no clear evidence for variation in the spectral index across the remnant; note that all regions agree at the two-sigma level or better, and the statistical error on the northern region remains quite large. However, the poor reduced $\chi^2$ of Region 1 indicates that a simple power law does not provide a good description of the spectrum in that region, possibly indicating a break or cut-off is present.

\begin{table}[htdp]
\caption{Results of power-law fits to spectral points extracted from several regions in the TeV emission from IC 443.}
\label{tab:spectra}
\begin{center}
\begin{tabular}{|r|c|c|c|c|}
\hline
Region & Position & Normalization & Index & $\chi^2$ / ndf \\
& & ($/550$ GeV) * 10$^{-13}$ TeV & & \\
\hline
Entire Remnant & 06 16 52.8 +23 33 00 & 9.92 $\pm$ 0.90 & -2.80 $\pm$ 0.09 & 2.76 / 3 \\
\hline
Region 1 & 06 16 43.2 +22 33 00 & 3.69 $\pm$  0.42 & -3.15 $\pm$ 0.11 & 9.98 / 3 \\
\hline
Region 2 & 06 17 52.8 +22 29 24 & 2.33 $\pm$ 0.42 & -3.19 $\pm$ 0.17 & 1.85 / 3 \\
\hline
Region 3 & 06 17 33.6 +22 45 36 & 1.86 $\pm$ 0.49 & -2.49 $\pm$ 0.42 & 2.64 / 3 \\
\hline
\end{tabular}
\end{center}
\label{default}
\end{table}%

\begin{figure}[h!]
\centering
\null\hfill
\subfloat{\includegraphics[width=0.5\textwidth]{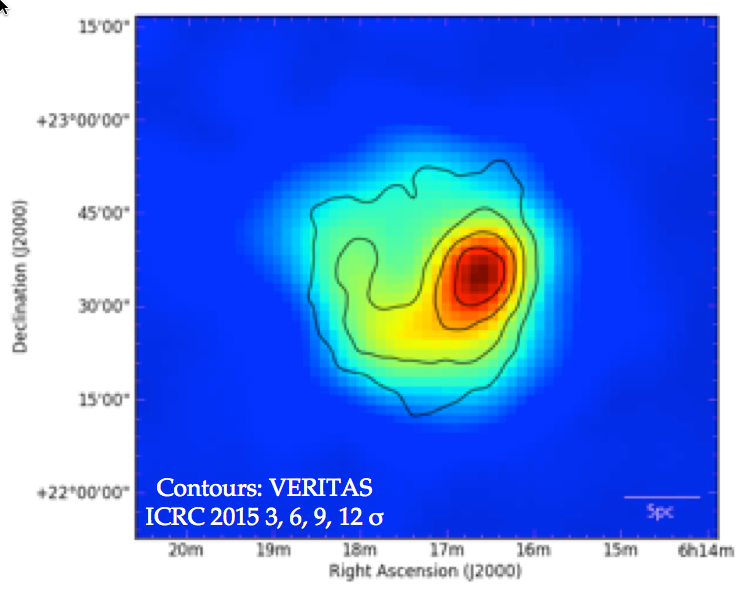}}
\hfill
\subfloat{\includegraphics[width=0.45\textwidth]{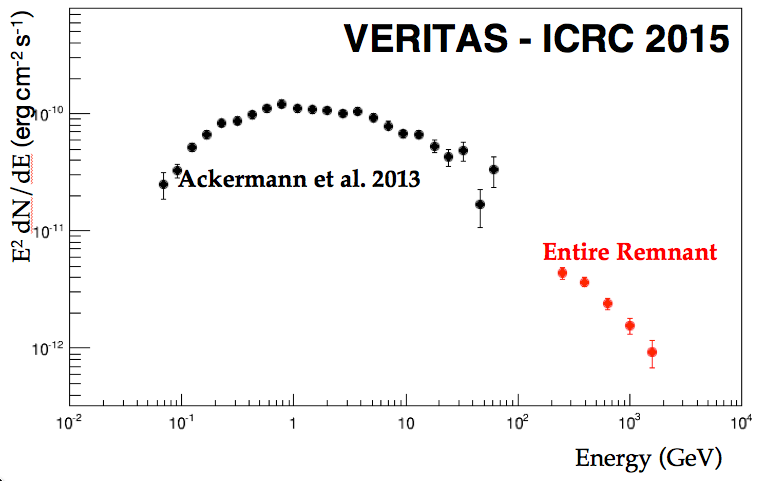}}
\hfill\null
\caption{(left) \emph{Fermi}-LAT counts map, with VERITAS significance contours overlaid. (right) Gamma-ray SED for the entire remnant.}
\label{fig:fermiVERITAS}
\end{figure}

Of great interest is the comparison of the VHE morphology with the GeV morphology, and Figure~\ref{fig:fermiVERITAS} (left) shows a counts map of \emph{Fermi}-LAT photons selected above $5$ GeV with VERITAS significance contours overlaid. The LAT counts map is derived from 83 months of data, using the P8R2\_SOURCE\_V6 response functions and Fermi Science Tools v10r0p5 for the analysis. The 50\% of events with the best PSF (event classes PSF2 and PSF3) have been chosen to provide an angular resolution comparable to VERITAS. Plotting the VERITAS and Fermi spectra together, as in Figure~\ref{fig:fermiVERITAS} (right), we see that the broadband SED also looks like a smooth continuation of the spectrum from the GeV to the VHE range. The remarkable degree of correlation in morphology and smooth continuation of the gamma-ray spectrum from the GeV range to the TeV range  is suggestive that the broad-band gamma-ray emission may be originating from a single population of cosmic rays interacting with shocked gas. If so, this would be in contrast to earlier models of the gamma-ray emission from IC 443 such as~\cite{Torres2010}, which argued for separate populations of CRs interacting with different MCs; in particular, the GeV emission arising from an MC within a few parsecs of the shock front, and the TeV emission from a second, larger cloud at a distance of 10-30 parsecs outside the shock front and interacting with a population of diffusing CRs.

\begin{figure}[h!]
\centering
\null\hfill
\subfloat{\includegraphics[width=0.38\textwidth]{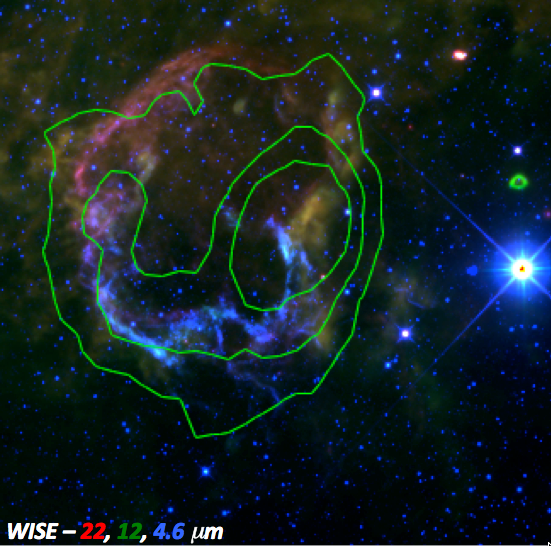}}
\hfill
\subfloat{\includegraphics[width=0.38\textwidth]{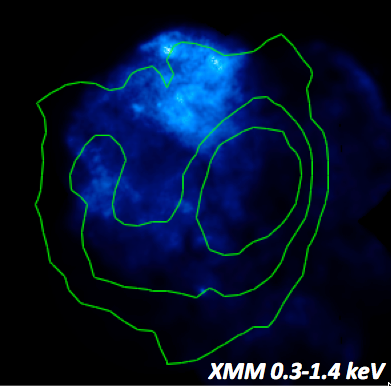}}
\hfill\null
\caption{\label{fig:wiseXMMVERITAS}(left) WISE image of IC 443 in 22 $\mu$m (red), 12 $\mu$m (green), and 4.6 $\mu$m (blue) channels, with VERITAS significance contours (3, 6, 9 $\sigma$) overlaid.  (right) XMM-Newton image of IC 443 in the $0.3 - 1.4\ \textrm{keV}$ band, with the same VERITAS contours overlaid.}
\end{figure}

\begin{figure}[h!]
\centering
\null\hfill
\subfloat{\includegraphics[width=0.38\textwidth]{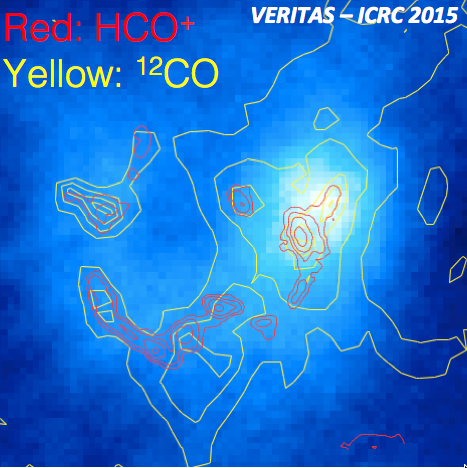}}
\hfill
\subfloat{\includegraphics[width=0.38\textwidth]{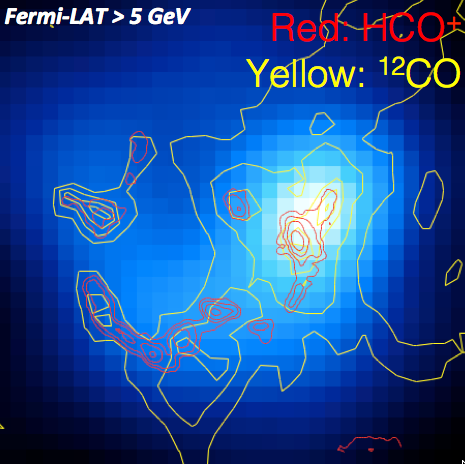}}
\hfill\null
\caption{\label{fig:gas}(left) VERITAS excess map with contours of two tracers of shocked gas, HCO$^+$ (red) and $^{12}$CO (yellow) overlaid.  (right) The \emph{Fermi}-LAT counts map, with the same contours overlaid.}
\end{figure}

Figures~\ref{fig:wiseXMMVERITAS} and \ref{fig:gas} compare the distributions of GeV and TeV emission to several tracers of the gas in the region within and surrounding IC 443. Figure~\ref{fig:wiseXMMVERITAS} (left) overlays the TeV contours on a multi-band image of IC 443 in the IR~\cite{Su2014}, revealing a strong correlation with dust emission in the region, particularly in the $4.6\ \mu\textrm{m}$ and $12\ \mu\textrm{m}$ ranges. Figure~\ref{fig:wiseXMMVERITAS} (right) shows that the TeV emission is anticorrelated with the thermal X-ray emission as seen by \emph{XMM}. This is most likely due to absorption of the X-rays by intervening dense gas along the lines of sight to the western and southern parts of the remnant.

Figure~\ref{fig:gas} (left) and (right) overlay contours of shocked gas traced by HCO$^+$ (red) and $^{12}$CO (yellow, $|v_{LSR}| < 10\ \textrm{km/s}$) on the VERITAS and \emph{Fermi}-LAT excess maps, respectively~\cite{Lee2012}. In both cases, the gamma-ray emission correlates strongly with the shocked gas, again arguing that the GeV and TeV emission arise from the same population of CRs. In that case, the emission is likely dominated by CRs interacting with gas close to the shock front, rather than an escaping population of CRs. 

\section{Conclusions}
Deep VERITAS observations of the supernova remnant IC 443 have established that its VHE gamma-ray emission is extended over the entire surface of the remnant and traces out the shell, thus adding IC 443 to the small but growing list of VHE shell-type SNRs. The morphology is strongly correlated with the GeV morphology, suggesting that the emission is dominated by a single population of CRs across a wide range of energies. IC 443 is the first VHE shell-type SNR to clearly have significant SNR/MC interactions, and is likely the oldest and most evolved of the VHE shell-type SNRs. As such, it remains an extremely interesting laboratory in which to study the acceleration, escape, and diffusion of cosmic rays.

\section*{Acknowledgements}
This research is supported by grants from the U.S. Department of Energy Office of Science, the U.S. National Science Foundation and the Smithsonian Institution, and by NSERC in Canada. We acknowledge the excellent work of the technical support staff at the Fred Lawrence Whipple Observatory and at the collaborating institutions in the construction and operation of the instrument. 

The VERITAS Collaboration is grateful to Trevor Weekes for his seminal contributions and leadership in the field of VHE gamma-ray astrophysics, which made this study possible.

\end{document}